\documentclass[12pt,a4paper]{article}
\usepackage{float}
\usepackage[normalem]{ulem}
\usepackage{subcaption}
\usepackage{gensymb}
\usepackage{amsmath}
\usepackage{wrapfig}
\usepackage{amssymb}
\usepackage{comment}
\usepackage{epsfig}
\usepackage{graphicx}
\usepackage{caption}
\usepackage[hidelinks]{hyperref}
\usepackage{csquotes}
\graphicspath{{plots/}}
\usepackage{multirow}
\usepackage{pdflscape}
\usepackage{xcolor}
\usepackage[title]{appendix}
\usepackage[left=.45in,right=.45in,top=.6in,bottom=.6in,headheight=14.5pt]{geometry}
\usepackage{fmtcount} 
\usepackage{array,multirow}
\newcolumntype{C}[1]{>{\centering\arraybackslash}m{#1}}
\usepackage{epstopdf}
\usepackage{graphicx}
\usepackage{geometry}
\usepackage{amsmath}
\usepackage{cite}
\usepackage{ctable}

\newgeometry{vmargin={10mm,20mm},hmargin={19mm,19mm}}
    \usepackage{adjustbox}
\usepackage{lipsum}

\begin{document}

\title{Neutrino Mass Matrix with broken Scaling in light of LMA and Dark-LMA Solutions}

\author{Ajay Kumar\thanks{ajaykumarksg72@gmail.com}, Dikshit Gautam\thanks{gautamrekha327@gmail.com} and Surender Verma\thanks{s\_7verma@hpcu.ac.in}}

\date{%
Department of Physics and Astronomical Science\\
Central University of Himachal Pradesh\\
Dharamshala, India-176215
}
\maketitle
\begin{abstract}
\noindent In the present work we have investigated some patterns of broken ``scaling" ansatz of the neutrino mass matrix. The scaling neutrino mass matrix is disallowed by the current neutrino oscillation data as, among others, it predicts vanishing reactor angle ($\theta_{13}=0$). We study its possible breaking scenarios in light of the large mixing angle (LMA) and Dark-LMA solutions suggested by current neutrino oscillation data. The normal hierarchical neutrino mass spectrum is ruled out in all three possible breaking patterns. Also, one of the interesting features of these breaking scenarios is the interplay between $\theta_{23}$-octant and possible CP violation. We find that the model allows maximal CP violation for $\theta_{23}$ above $6\%$ of its maximal value which, interestingly, is close to its current best-fit value for inverted hierarchical neutrino masses. We have, also, investigated the implications for effective Majorana neutrino mass parameter $|M_{ee}|$ for allowed breaking patterns.  The correlation behavior of Majorana CP phases, which can be probed in $0\nu\beta\beta$ decay experiments, is found to have the capability of distinguishing LMA and Dark-LMA solutions.

\end{abstract}

\section{Introduction}
The Standard Model (SM) of particle physics, while remarkably successful in describing fundamental particles and their interactions, has several limitations. It does not provide a mechanism for neutrino masses, which are now firmly established by oscillation experiments. Additionally, the SM fails to incorporate dark matter and dark energy, which constitute most of the universe's mass-energy content. Various new physics scenarios have been considered to explain small neutrino masses. Amongst them are the well known frameworks based on UV completions of the dimension-5 Weinberg operator at tree level called seesaw mechanism\cite{Minkowski:1977sc, Yanagida:1979as, Gell-Mann:1979vob, Mohapatra:1979ia}. An extension of the SM, generally, lead to Yukawa matrices for charged leptons and neutrinos after spontaneous breaking of the gauge symmetry, which oftenly, is augmented by an additional flavor symmetry based on some discrete or continuous group. For recent progress in model building approaches based on seesaw mechanisms, radiative mass generation models, and models incorporating discrete symmetries, see Refs. \cite{Verma:2015tpa,deGouvea:2016qpx}. The information about respective mass spectrum and mixing of the lepton can be extracted after diagonalization of the charged and neutrino mass matrices. The lepton mixing matrix, in general, is given by $V=V_l^\dagger V_\nu$, where $V_l$ and $V_\nu$ are the diagonalizing unitary matrices for charged lepton and neutrinos, respectively. In the charged lepton basis i.e., the basis in which charged lepton mass matrix is diagonal ($V_l$ is identity matrix), $V=V_\nu$. The matrix $V$ can be parametrized in terms of three mixing angles ($\theta_{12}, \theta_{13}, \theta_{23}$) and one (three) CP violating phase (s) considering neutrino to be of Dirac (Majorana) nature. The neutrino mass matrix, thus, contain, all the information about the underlying symmetry and is function of nine physical observables constituting three neutrino mass eigenvalues ($m_1,m_2,m_3$), three mixing angles ($\theta_{12}, \theta_{13}, \theta_{23}$) and three CP violating phases ($\delta,\alpha,\beta$) where $\delta$ ($\alpha,\beta$) is (are) Dirac (Majorana)-type CP violating phase(s). The complete reconstruction of the neutrino mass matrix, is not possible as we do not have information on the absolute neutrino mass scale and CP violating phases. In view of this, the phenomenological approaches  play a crucial role in understanding neutrino properties, interactions, and potential new physics. The imposition of texture zero(s)\cite{Frampton:2002yf,Desai:2002sz,Dev:2006qe,Dev:2008cj,Lashin:2011dn,Dev:2012xn,Verma:2020gpl,Singh:2022ijf,Raj:2024dph,Chamoun:2023vnn}, hybrid texture\cite{Frigerio:2002fb,Kaneko:2005yz,Dev:2009he,Dev:2010vy,Singh:2018gvw}, vanishing determinant condition\cite{Branco:2002ie,Chauhan:2006uf,Singh:2018lao} and vanishing minor\cite{Lashin:2009yd,Dev:2010if,Wang:2013woa,Dey:2023tsk}, to name a few, are some examples of such phenomenological ansatze. The important consequence of these approaches is the emergence of correlations among the parameters which can be probed in experiments. For example, current/future neutrinoless double beta ($0\nu\beta\beta$) decay experiments may shed light of the yet unknown Majorana-type CP violating phases and absolute scale of neutrino mass. 

\noindent Another phenomenological structure of neutrino mass matrix having third column scaled with respect to second column called ``Scaling" ansatz has, also, been proposed\cite{Mohapatra:2006xy}. A scaled neutrino mass matrix is of rank 2 which predicts vanishing neutrino mass eigenvalue ($m_3=0$), $\theta_{13}=0$ and a maximal value of $\theta_{23}$ thus, rendering scaling ansatz disallowed by current neutrino oscillation data\cite{Esteban:2024eli}. Various possibilities of its breaking or deviations have, also, been investigated to ascertain its viability in light of the standard three generation large mixing angle (LMA) solution\cite{Blum:2007qm,Joshipura:2009fu, Rodejohann:2010zza, Verma:2012ed,Dev:2013esa,Kalita:2014vxa,Chakraborty:2014hfa, Ghosal:2015lwa,Samanta:2016nat,Sinha:2017rjj,Kashav:2021zir,Nagao:2024yoe}. 

\noindent However, any possible extension of the SM to include small neutrino mass may give rise to non-standard interactions (NSIs) of neutrino with matter\cite{Wolfenstein:1977ue}. In presence of NSIs, one get new solution space consistent with neutrino oscillation data with $\sin^2 \theta_{12}\simeq 0.7$\cite{Miranda:2004nb,Escrihuela:2009up}. This new solution space is called Dark-LMA (DLMA) and has important implications from point of views of associated phenomenology and model building approaches as well. In the present work, we study the phenomenology of broken scaling in light of LMA and Dark-LMA solutions.

\noindent The outline of this paper is as follows. In Section \ref{sec2}, we introduce the framework and
discuss the constraining equations emanating from three possible scale breaking scenarios. In Section \ref{sec3} we explain numerical analysis and discuss predictions of the model. Finally, we summarize in Section \ref{sec4}.

\section{Formalism}\label{sec2}
The scaling ansatz of Majorana neutrino mass matrix ($M_\nu$) has been obtained using Type-I seesaw mechanism viz.,
  \begin{equation}\label{eq.1}
M_\nu =M_{D}M^{-1}_{R} M^{T}_{D}=\begin{pmatrix}
a ~& ~b ~& ~\frac{b}{s} \\
b ~&~ d~ & ~\frac{d}{s} \\
\frac{b}{s}~ & ~\frac{d}{s}~ & ~\frac{d}{s^2}\\
\end{pmatrix},
\end{equation}
where $M_D$ and $M_R$ are Dirac and right-handed Majorana neutrino mass matrices, respectively. The structure of $M_\nu$ (Eqn. (\ref{eq.1})) necessarily occurs independent of the form of $M_R$, if third row of $M_D$ multiplied with $s$ is equal to the second
row\cite{Joshipura:2009fu}. $M_\nu$ is of rank 2 with  ratio of its elements being equal, which is termed as the Strong Scaling Ansatz (\text{SSA})\cite{Mohapatra:2006xy}.
The complex symmetric neutrino mass matrix, in charged lepton basis, is given by 
\begin{equation}\label{eq.2}
   M_\nu =\begin{pmatrix}
M_{ee} & M_{e\mu} & M_{e\tau} \\
M_{e\mu} &  M_{\mu \mu} & M_{\mu \tau} \\
M_{e\tau} &  M_{\mu \tau}  & M_{\tau \tau}\\
\end{pmatrix}= V^*m_\nu^d V^\dagger,
\end{equation}
where $m_\nu^d$ is diagonal neutrino mass eigenvalue matrix, which is given by $diag(m_1,m_2 ,m_3)$ and  $V=U.P$ is complex unitary matrix
\begin{align}\label{7}
    U =&\begin{pmatrix}
U_{e1} &U_{e2}& U_{e3} \\
U_{\mu1}&
U_{\mu2}&U_{\mu3} \\
U_{\tau1}& U_{\tau2}& U_{\tau3}\end{pmatrix}=&\begin{pmatrix}
c_{12} c_{13} & c_{13} s_{12}& e^{-\iota \delta} s_{13} \\
-c_{23} s_{12} - c_{12} e^{\iota \delta} s_{13} s_{23}&
c_{12} c_{23} - e^{\iota \delta} s_{12} s_{13} s_{23}&c_{13} s_{23} \\
-c_{12} c_{23} e^{\iota \delta} s_{13} + s_{12} s_{23}& -c_{23} e^{\iota \delta}s_{12} s_{13} - 
 c_{12} s_{23}& c_{13} c_{23}\\
\end{pmatrix},
\end{align}
where matrix $U$ is Pontecorvo-Maki-Nakagawa-Sakata (\text{PMNS}) matrix with $c_{ij}=\cos\theta_{ij}$ and $s_{ij}=\sin\theta_{ij}$ and the Dirac type \text{CP}-violating phase $\delta$. $P$ is the phase matrix, which includes Majorana phases $\alpha$ and $\beta$, given by
\begin{align}
P=
&\begin{pmatrix}
1 & 0 &0 \\
0&  e^{\iota\alpha} &0 \\
0 &  0  &e^{\iota(\beta+\delta)} \\
\end{pmatrix}.
\end{align}  
The effective Majorana neutrino mass matrix can be written as
\begin{equation}\label{6}
      M_\nu
= U^*\tilde{m}_\nu^d U^\dagger,
\end{equation}
where $\tilde{m}_\nu^d= \text{diag}(\lambda_1,\lambda_{2} ,\lambda_{2})$ with 
\[\lambda_{1}=m_{1},~\lambda_2=m_2e^{-2\iota\alpha},~\lambda_3=m_3e^{-2\iota(\beta+\delta)}.\]
The scaling structure can be translated as
\begin{equation}
  \frac{M_{e\mu}}{M_{e\tau}}=   \frac{M_{\mu\mu}}{M_{\mu\tau}}=  \frac{M_{\tau\mu}(=M_{\mu\tau})}{M_{\tau \tau}}=s,
\end{equation}
which results in three conditions viz.,
\begin{align}
  &{M_{e\mu}}-s{M_{e\tau}}=0,\label{3} \\
  &{M_{\mu\mu}}-s{M_{\mu\tau}}=0,\label{4}\\
  &{M_{\mu\tau}}-s{M_{\tau \tau}}=0\label{5}, 
\end{align}
where $s$ is the ``scaling factor".
However, the problem with scaling is that it predicts  vanishing neutrino mass ($m_3=0$) and predicts no \text{CP}-violation in neutrino oscillation experiments. So, the scale breaking parameter $\epsilon$ is introduced in Eqns. (\ref{3}), (\ref{4}) and (\ref{5}), leading to three cases each with three constraint equations, shown in  Table \ref{taba}.
 \begin{table}[t]

\begin{center}
\resizebox{0.6\textwidth}{!}{%
\begin{tabular}{ccl}
\hline
Scale breaking Case&Mass matrix& Constraint Equations\\
\hline\hline\\
$b_1$&$\begin{pmatrix}
a & b & \frac{b}{s(1+\epsilon)} \\
b & d & \frac{d}{s} \\
\frac{b}{s} & \frac{d}{s} & \frac{d}{s^2}\\
\end{pmatrix}$&
$\begin{array}{l}
  {M_{e\mu}}-s(1+\epsilon){M_{e\tau}}=0 \\
  {M_{\mu\mu}}-s{M_{\mu\tau}}=0\\
  {M_{\mu\tau}}-s{M_{\tau \tau}}=0\\  
\end{array}$\\
\\
 \hline
 \\
$b_2$& $\begin{pmatrix}
a & b & \frac{b}{s} \\
b & d & \frac{d}{s(1+\epsilon)} \\
\frac{b}{s} & \frac{d}{s} & \frac{d}{s^2}\\
\end{pmatrix}$&
$\begin{array}{l}
  {M_{e\mu}}-s{M_{e\tau}}=0 \\
  {M_{\mu\mu}}-s(1+\epsilon){M_{\mu\tau}}=0\\
  {M_{\mu\tau}}-s{M_{\tau \tau}}=0\\  
\end{array}$\\
\\
  \hline
  \\
$b_3$& $\begin{pmatrix}
a & b & \frac{b}{s} \\
b & d & \frac{d}{s} \\
\frac{b}{s} & \frac{d}{s} & \frac{d}{s^2(1+\epsilon)}\\
\end{pmatrix}$&
$\begin{array}{l}
  {M_{e\mu}}-s{M_{e\tau}}=0 \\
  {M_{\mu\mu}}-s{M_{\mu\tau}}=0\\
  {M_{\mu\tau}}-s(1+\epsilon){M_{\tau \tau}}=0\\  
\end{array}$\\
\\
 \hline
 \end{tabular}%
 }
\caption{\label{taba} The three possibilities of broken scaling in $M_\nu$ and corresponding equations  where $a,b,d$ are complex and $s$, $\epsilon$ are real parameters.}
\end{center}
\end{table}%
Here $b_1,b_2$ and $ b_3 $ corresponds to the breaking in first, second and third row of the effective Majorana neutrino mass matrix (Eqn. (\ref{eq.1})), respectively.
Furthermore, these equations can be written in terms of $\lambda_1,\lambda_2$ and $\lambda_3$ using Eqns. (\ref{7}) and (\ref{6}), shown in Table \ref{tabb}. 

\noindent These constraints equations carries more than three unknowns. So, we write these in terms of mass ratios $r_{21}\equiv\frac{\lambda_2}{\lambda_1}$ and $r_{21}\equiv\frac{\lambda_3}{\lambda_1}$.
In general,
\begin{align}
   &  a_1+r_{21}a_2+r_{31}a_3=0, \\
    &  a'_1+r_{21}a'_2+r_{31}a'_3=0,\label{8}\\
     & a''_1+r_{21}a''_2+r_{31}a''_3=0,\label{9} 
\end{align}
where the coefficients ($a_1,a_2,a_3$), ($a'_1,a'_2,a'_3$) and ($a''_1,a''_2,a''_3$) can be easily inferred from second and third columns of Table \ref{tabb}, for different broken scenarios. In each case, we solve first two equations (excluding equation with breaking parameter $\epsilon$) to get two mass ratios ($\lambda_1, \lambda_2$) and the Majorana phases $\alpha$ and $\beta$ which are shown in Table \ref{69}.

 \begin{table}[H]

\begin{center}

\resizebox{0.8\textwidth}{!}{%
\begin{tabular}{clc}
\hline
Scale breaking case& Constraint Equations& Variables\\
\hline\hline\\
$b_1$&$\begin{array}{l}
      x\lambda_1+y\lambda_2+z\lambda_3=0 \\
      x'\lambda_1+y'\lambda_2+z'\lambda_3=0\\
      x''\lambda_1+y''\lambda_2+z''\lambda_3=0
\end{array}$&
$\begin{array}{l}
x=U_{e1}( U_{\mu1}^* - (1 +\epsilon) s U_{\tau1}^*)\\
y=U_{e2}( U_{\mu2}^* - (1 +\epsilon) s U_{\tau2}^*)\\
z=U_{e3}^*( U_{\mu3} - (1 +\epsilon) s U_{\tau3})\\
  x'=U_{\mu1}^* (U_{\mu1}^* - s U_{\tau1}^*) \\
y'=U_{\mu2}^* (U_{\mu2}^* - s U_{\tau2}^*) \\
z'=U_{\mu3} (U_{\mu3} - s U_{\tau3}) \\
x''=U_{\tau1}^* (U_{\mu1}^* - s U_{\tau1}^*)\\
y''=U_{\tau2}^* (U_{\mu2}^* - s U_{\tau2}^*)\\
y''=U_{\tau3} (U_{\mu3} - s U_{\tau3})\\
\end{array}$\\
\\
 \hline
 \\
$b_2$& $\begin{array}{l}
      u\lambda_1+v\lambda_2+w\lambda_3=0 \\
      r\lambda_1+s\lambda_2+t\lambda_3=0\\
      x''\lambda_1+y''\lambda_2+z''\lambda_3=0
\end{array}$&
$\begin{array}{l}
 u=U_{e1} (U_{\mu1}^* - sU_{\tau1}^*)\\
 v=U_{e2} (U_{\mu2}^* - sU_{\tau2}^*)\\
  w=U_{e3}^* (U_{\mu3} - sU_{\tau3})\\
  r=U_{\mu1}^*(U_{\mu1}^* - (1 + \epsilon) sU_{\tau1}^*)\\
   s=U_{\mu2}^*(U_{\mu2}^* - (1 + \epsilon) sU_{\tau2}^*)\\
    t=U_{\mu3}(U_{\mu3} - (1 + \epsilon) sU_{\tau3})\\
  
\end{array}$\\
\\
  \hline
  \\
$b_3$& $\begin{array}{l}
      u\lambda_1+v\lambda_2+w\lambda_3=0 \\
      x'\lambda_1+y'\lambda_2+z'\lambda_3=0\\
      x'''\lambda_1+y'''\lambda_2+z'''\lambda_3=0
\end{array}$&
$\begin{array}{l}
x'''= U_{\tau1}^*(U_{\mu1}^*  - s  U_{\tau1}^* (1 + \epsilon)) \\
 y'''= U_{\tau2}^*(U_{\mu2}^*  - s  U_{\tau2}^* (1 + \epsilon)) \\
 z'''= U_{\tau3}(U_{\mu3}  - s  U_{\tau3} (1 + \epsilon)) \\
\end{array}$\\
\\
 \hline
 \end{tabular}%
 }
\caption{\label{tabb}The constraining equations for each scale breaking cases $b_1,b_2$ and $b_3$ in terms of $\lambda_1,\lambda_2,\lambda_3$ and elements of $U$.}
\end{center}
\end{table}

\noindent Using mass ratios and Majorana phases we look for allowed parameter space for which third constraining equation (Eqn. (\ref{9})) is satisfied. From the Neutrino oscillation data, we know two mass-squared differences $\Delta m^2_{21}=m_2^2-m_1^2$ and $|\Delta m^2_{32}|=|m_3^2-m_2^2|$, which lead to two possible mass hierarchy for neutrinos. Using mass ratios $r_{21}$ and $r_{31}$ yield two values of neutrino mass $m_1$ given by
\begin{align}
&m_{1a}=\sqrt{\frac{\Delta m_{21}^2}{|r_{21}|^2-1}},\\~~ &m_{1b}=\sqrt{\frac{\Delta m_{32}^2}{|r_{31}|^2-|r_{21}|^2}}, ~~\text{for normal hierarchy (NH)}~(m_3>m_2>m_1),  \\
&m_{1b}=\sqrt{\frac{\Delta m_{32}^2}{|r_{31}|^2-|r_{21}|^2}},~~\text{for inverted hierarchy (IH)}~(m_2>m_1>m_3),
\end{align}
respectively.
For solution to be self-consistent these two values $(m_{1a},m_{1b})$ must be equal. This condition can be translated to \\
\begin{equation}\label{rnu}
\frac{\Delta m_{21}^2}{\Delta m_{32}^2}=\frac{|r_{21}|^2-1}{|r_{31}|^2-|r_{21}|^2},
\end{equation}
where we define parameter $R_\nu\equiv\frac{\Delta m_{21}^2}{\left|\Delta m_{32}^2\right|}$. The $3\sigma$ experimental range of this parameter is $0.026 < R_\nu < 0.037$.  So, the allowed phenomenology of the model is obtained by invoking the condition that $\left|\frac{m_{1a}-m_{1b}}{m_{1a}}\right|$ should be vanishingly small with in the defined tolerance of $\approx \mathcal{O}(10^{-2})$.

\noindent The neutrino mass eigenstates $m_2$ and $m_3$ can then be obtained using the mass-squared differences as
\begin{align}
    &m_2=\sqrt{\Delta m_{21}^2+m_{1}^2},\\
    &m_3=\sqrt{\Delta m_{32}^2+m_{2}^2},~~~ \text{for NH},\\
    &m_3=\sqrt{m_{1}^2-\Delta m_{23}^2+\Delta m_{21}^2},~~~ \text{for IH}\label{m3s}.
\end{align}
For $m_3$ to be real one requires $m_{1}^2-\Delta m_{23}^2+\Delta m_{21}^2$, in Eqn. (\ref{m3s}), must be greater than 0. While performing the numerical analysis for LMA scenario, all the known parameters such as mixing angles $\theta _{ij}~(i,j=1,2,3;i<j)$ (except $\theta_{23}$) and mass-squared differences $\Delta m_{ij}^2 ~(i>j)$ are randomly generated using Gaussian distribution with central values and errors given in Table \ref{data}. As octant of $\theta_{23}$ is still unknown, so we want to study the implication of the model for $\theta_{23}$-octant, thus, we have set it free, without any initial bias, and varied it in $3\sigma$ range using uniform distribution  allowing values above as well as below maximality with equal probability.
The Dirac phase ($\delta$) and the free parameters $s$ and $\epsilon$  are randomly generated in the ranges $\delta\in (0^o,360^o)$, $s\in (-1.5,-0.5)$ and $\epsilon\in (1,2)$ using uniform distributions. 

\begin{table}[t]

\begin{center}
\resizebox{1.08\textwidth}{!}{%
\begin{tabular}{ccc}

\hline
Scale breaking &Absolute Mass Ratios& Majorana Phases\\
\hline\hline\\
$b_1$&$\begin{array}{l}
    \left|\frac{\lambda_2}{\lambda_1}\right|=\frac{m_2}{m_1}= \left|\frac{(U_{e3}^* U_{\mu1}^* - U_{e1} U_{\mu3}) (U_{\mu1}^* - s U_{\tau1}^*)}{(U_{e3}^* U_{\mu2}^* - 
    U_{e2} U_{\mu3}^*) (-U_{\mu2}^* + s U_{\tau2}^*)}\right| \\ 
 \\
     \left|\frac{\lambda_3}{\lambda_1}\right| =\frac{m_3}{m_2}=\left|\frac{(U_{e2} U_{\mu1}^* - U_{e1} U_{\mu2}^*) (U_{\mu1}^* - s U_{\tau1}^*)}{(U_{e3}^* U_{\mu2}^* - 
    U_{e2} U_{\mu3}^*) (U_{\mu3} - s U_{\tau3})}\right|
      
\end{array}$&
$\begin{array}{l}
\alpha= -\frac{1}{2} \text{Arg}\left( \frac{(U_{e3}^* U_{\mu1}^* - U_{e1} U_{\mu3}) (U_{\mu1}^* - s U_{\tau1}^*)}{(U_{e3}^* U_{\mu2}^* - 
    U_{e2} U_{\mu3}^*) (-U_{\mu2}^* + s U_{\tau2}^*)}\right) \\
   \\
   \beta=-\frac{1}{2} \text{Arg}\left(  \frac{(U_{e2} U_{\mu1}^* - U_{e1} U_{\mu2}^*) (U_{\mu1}^* - s U_{\tau1}^*)}{(U_{e3}^* U_{\mu2}^* - 
    U_{e2} U_{\mu3}^*) (U_{\mu3} - s U_{\tau3})}\right)-\delta 
\end{array}$\\
\\
 \hline
 \\
$b_2$& $\begin{array}{l}
    \left|\frac{\lambda_2}{\lambda_1}\right|= \frac{m_2}{m_1}= \left|\frac{  -(U_{\mu1}^* - s U_{\tau1}^*) (U_{e3}^* U_{\tau1}^* - U_{e1} U_{\tau3})}{(U_{\mu2}^* - 
    s U_{\tau2}^*) (U_{e3}^* U_{\tau2}^* - U_{e2} U_{\tau3})}\right|\\
     \\  \left|\frac{\lambda_3}{\lambda_1}\right|=\frac{m_3}{m_1}=   \left|\frac{  -(U_{\mu1}^* - s U_{\tau1}^*) (U_{e2} U_{\tau1}^* - U_{e1} U_{\tau2}^*)}{(U_{\mu3} - 
    s U_{\tau3}) (-U_{e3}^* U_{\tau2}^* + U_{e2} U_{\tau3})} \right|
      
\end{array}$&
$\begin{array}{l}
\alpha= -\frac{1}{2} \text{Arg}\left(\frac{  -(U_{\mu1}^* - s U_{\tau1}^*) (U_{e3}^* U_{\tau1}^* - U_{e1} U_{\tau3})}{(U_{\mu2}^* - 
    s U_{\tau2}^*) (U_{e3}^* U_{\tau2}^* - U_{e2} U_{\tau3})}\right)\\
  \\ \beta=-\frac{1}{2} \text{Arg}\left( \frac{  -(U_{\mu1}^* - s U_{\tau1}^*) (U_{e2} U_{\tau1}^* - U_{e1} U_{\tau2}^*)}{(U_{\mu3} - 
    s U_{\tau3}) (-U_{e3}^* U_{\tau2}^* + U_{e2} U_{\tau3})}\right)-\delta 
\end{array}$\\
\\
  \hline
  \\
$b_3$& $\begin{array}{l}
   \left|\frac{\lambda_2}{\lambda_1}\right|=  \frac{m_2}{m_1}=\left| \frac{ -(U_{\mu1}^* - s U_{\tau1}^*) (-U_{\mu3} U_{\tau1}^* + U_{\mu1}^* U_{\tau3})}{(U_{\mu2}^* - 
    s U_{\tau2}^*) (-U_{\mu3} U_{\tau2}^* + U_{\mu2}^* U_{\tau3})}\right| \\
    \\\left|\frac{\lambda_3}{\lambda_1}\right|= \frac{m_3}{m_1}=  \left|\frac{ -(U_{\mu1}^* - s U_{\tau1}^*) (-U_{\mu2}^* U_{\tau1}^* + U_{\mu1}^* U_{\tau2}^*)}{(U_{\mu3} - 
    s U_{\tau3}) (U_{\mu3} U_{\tau2}^* - U_{\mu2}^* U_{\tau3})}\right|
      
\end{array}$&
$\begin{array}{l}
\alpha= -\frac{1}{2} \text{Arg}\left(\frac{ -(U_{\mu1}^* - s U_{\tau1}^*) (-U_{\mu3} U_{\tau1}^* + U_{\mu1}^* U_{\tau3})}{(U_{\mu2}^* - 
    s U_{\tau2}^*) (-U_{\mu3} U_{\tau2}^* + U_{\mu2}^* U_{\tau3})}\right)\\
    \\ \beta=-\frac{1}{2} \text{Arg}\left( \frac{ -(U_{\mu1}^* - s U_{\tau1}^*) (-U_{\mu2}^* U_{\tau1}^* + U_{\mu1}^* U_{\tau2}^*)}{(U_{\mu3} - 
    s U_{\tau3}) (U_{\mu3} U_{\tau2}^* - U_{\mu2}^* U_{\tau3})}\right)-\delta 
\end{array}$\\
\\
 \hline
 \end{tabular}%
 }
\caption{\label{69} The mass ratios and Majorana phases $\alpha$ and $\beta$ for $b_1,b_2$ and $b_3$ cases.}
\end{center}
\end{table}

 \noindent However, for analysis of the Dark-LMA case, in all  three broken scenarios, $\theta _{12}$ is randomly generated in the range $\theta_{12}\in (53.71^o-58.37^o)$\cite{Escrihuela:2009up,Miranda:2007zza,Vishnudath:2019eiu,Farzan:2017xzy} using uniform distribution. The remaining details are same as that for LMA case.
 
\noindent Also, the rate of neutrinoless double beta decay $(0\nu \beta \beta)$ process is proportional to the modulus of the $M_{ee}$ element of the effective Majorana neutrino mass matrix  which is given as
\begin{align}
    \left|M_{ee}\right|&=\left|\sum_{i=1}^{3} \lambda_i {U_{ei}^{*}}^{2}\right| \nonumber\\
    &=\left|c_{12}^2 c_{13}^2 m_1 + e^{-2\iota\alpha}c_{13}^2 m_2 s_{12}^2 + e^{-2\iota \beta} m_3 s_{13}^2\right|.
\end{align}
The Jarlskog CP invariant which determines the magnitude of the CP violation in the neutrino oscillation is defined as
\begin{equation}
   J_{CP}=c_{12}s_{12}c_{13}^2s_{13}c_{23}s_{23}\sin{\delta}  .
\end{equation}

\begin{table}[t]\label{data}
\begin{center}

\begin{tabular}{l|l|l}
\hline
Parameter & Best fit $\pm$ \( 1 \sigma \) range & \( 3 \sigma \) range  \\
\hline\hline \multicolumn{2}{c} { Normal Hierarchy (NH) \( \left(m_{1}<m_{2}<m_{3}\right) \)} \\
\hline $\sin ^{2} \theta_{12} $ &  $0.308^{+0.012}_{-0.011} $ &  $0.275\rightarrow0.345$ \\
$\theta_{12}/\degree $ & $33.68^{+0.73}_{-0.70} $ &  $31.63 \rightarrow 35.95$ \\
$\sin ^{2} \theta_{13} $ & $0.02215^{+0.00056}_{-0.00058}$  & $0.02030\rightarrow 0.02388$ \\
$ \theta_{13}/\degree $ & $8.56_{-0.11}^{+0.11}$  & $8.19 \rightarrow 8.89$ \\
$\sin ^{2} \theta_{23} $ &$0.470^{+0.017}_{-0.013}$ & $0.435 \rightarrow 0.585$ \\
$ \theta_{23}/\degree $ & $43.3^{+1.0}_{-0.8}$ & $41.3 \rightarrow 49.9$ \\
$\delta_{CP}/\degree $ & $ 212^{+26}_{-41} $ & $124 \rightarrow 364$\\
\(\frac{ \Delta m_{21}^{2}}{10^{-5} ~\mathrm{eV}^{2}} \) & $7.49_{-0.19}^{+0.19}$& \(6.92 \rightarrow8.05 \) \\
\( \frac{\Delta m_{31}^{2}}{10^{-3}~ \mathrm{eV}^{2}}\) & $2.513_{-0.019}^{+0.021}$ & \(+2.451 \rightarrow 2.578 \) \\
\hline \multicolumn{2}{c} {Inverted Hierarchy (IH) \( \left(m_{3}<m_{1}<m_{2}\right) \)} \\
\hline $\sin ^{2} \theta_{12} $ &  $0.308^{+0.012}_{-0.011} $ &  $0.275\rightarrow0.345$ \\
$\theta_{12}/\degree $ & $33.68^{+0.73}_{-0.70} $ &  $31.63 \rightarrow 35.95$ \\
$\sin ^{2} \theta_{13} $ & $0.02231^{+0.00056}_{-0.00056}$  & $0.02060\rightarrow 0.02409$ \\
$ \theta_{13}/\degree $ & $8.59_{-0.11}^{+0.11}$  & $8.25 \rightarrow 8.93$ \\
$\sin ^{2} \theta_{23} $ &$0.550^{+0.012}_{-0.015}$ & $0.440 \rightarrow 0.584$ \\
$ \theta_{23}/\degree $ & $47.9^{+0.7}_{-0.9}$ & $41.5 \rightarrow 49.8$ \\
$\delta_{CP}/\degree $ & $ 274^{+22}_{-25} $ & $201 \rightarrow 335$\\
\(\frac{ \Delta m_{21}^{2}}{10^{-5} ~\mathrm{eV}^{2}} \) & $7.49_{-0.19}^{+0.19}$& \(6.92 \rightarrow8.05 \) \\
\( \frac{\Delta m_{32}^{2}}{10^{-3}~ \mathrm{eV}^{2}}\) & $-2.484_{-0.020}^{+0.020}$ & \(-2.547 \rightarrow -2.421 \) \\
\hline 
\end{tabular}
\caption{\label{data} The values of neutrino mixing angles and mass-squared differences used in the numerical analysis\cite{Esteban:2024eli}.}
\end{center}
\end{table}

\section{Numerical Analysis and Discussion}\label{sec3}
In this section, we discuss the result of the scale breaking $b_1,b_2$ and $b_3$ under the framework of LMA and Dark-LMA solutions. We have obtained the allowed parameter space for each scale breaking case which is consistent under LMA and Dark-LMA solutions. 
\subsection{$b_1$-Case}
In order to comprehend the numerical results we give the Taylor series expansion of the mass ratios up to the first order in $s_{13}$ viz.,
\begin{eqnarray}
&&|r_{21}|=\frac{m_2}{m_1} \approx \frac{s_{12}^2}{c_{12}^2} - \frac{s_{12}(s c_{23}-s_{23})}{c_{13}^3 (c_{23}+s s_{23})} s_{13}\cos\delta,\\
&&|r_{31}|=\frac{m_3}{m_1}\approx \frac{s_{12}(c_{23}+ s s_{23})}{c_{12}(s_{23}-s c_{23})} s_{13}.
\end{eqnarray}

\begin{figure}[t]
  \centering
\begin{tabular}{cc} 
\includegraphics[width=0.45\linewidth]{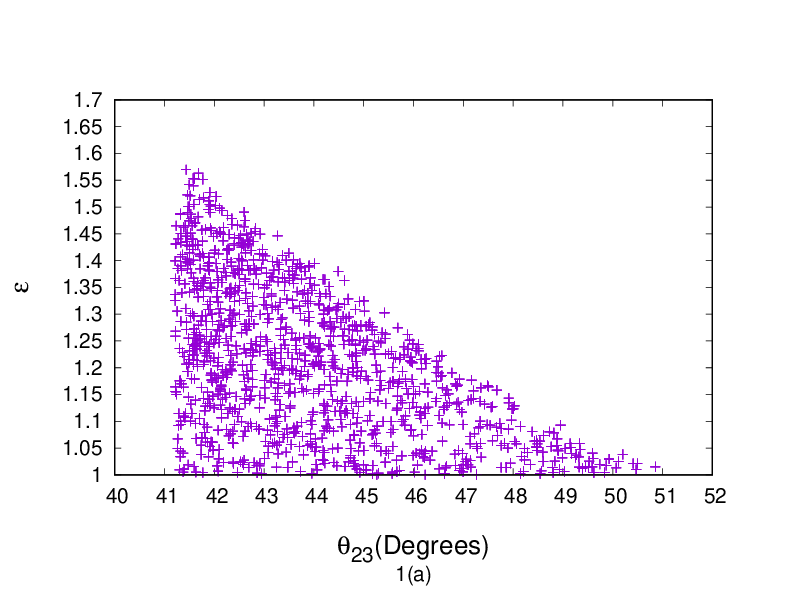}& \includegraphics[width=0.45\linewidth]{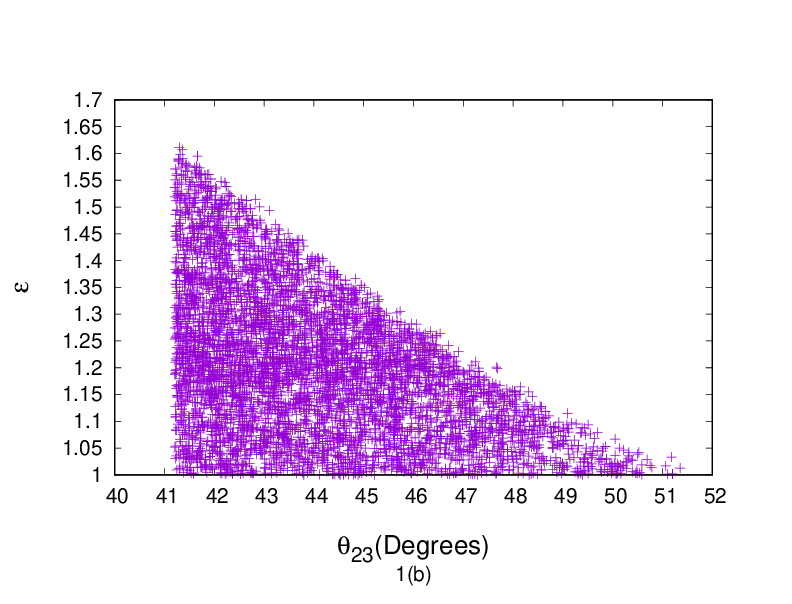}\\ 
 \includegraphics[width=0.45\linewidth]{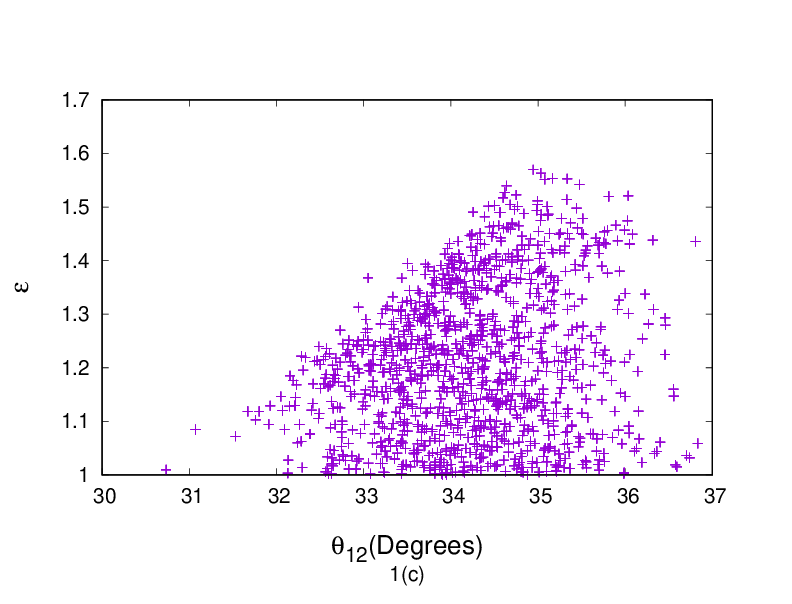}& \includegraphics[width=0.45\linewidth]{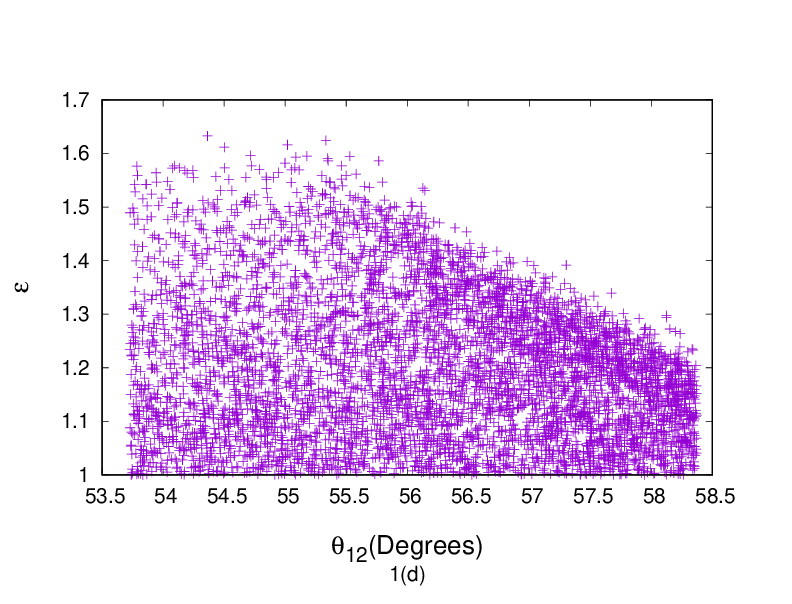}\\ 
\end{tabular}
  \caption{\textbf{$b_1$-Case:} The correlation plots between mixing angles $\theta_{12}$ and $\theta_{23}$ with scale breaking parameter $\epsilon$ with IH under LMA (left panel) and Dark-LMA (right panel) solutions.}
  \label{fig1}
\end{figure}

\begin{figure}[H]
  \centering
\begin{tabular}{cc} 
\includegraphics[width=0.45\linewidth]{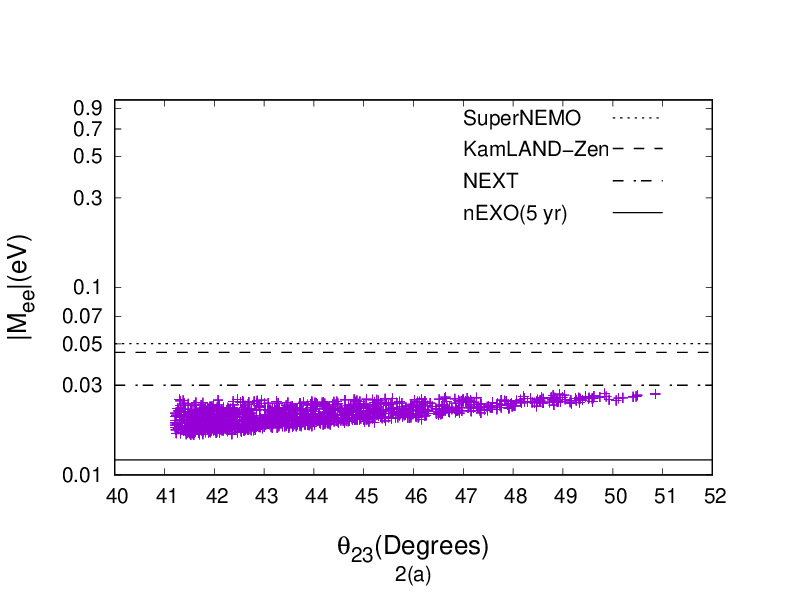}& \includegraphics[width=0.45\linewidth]{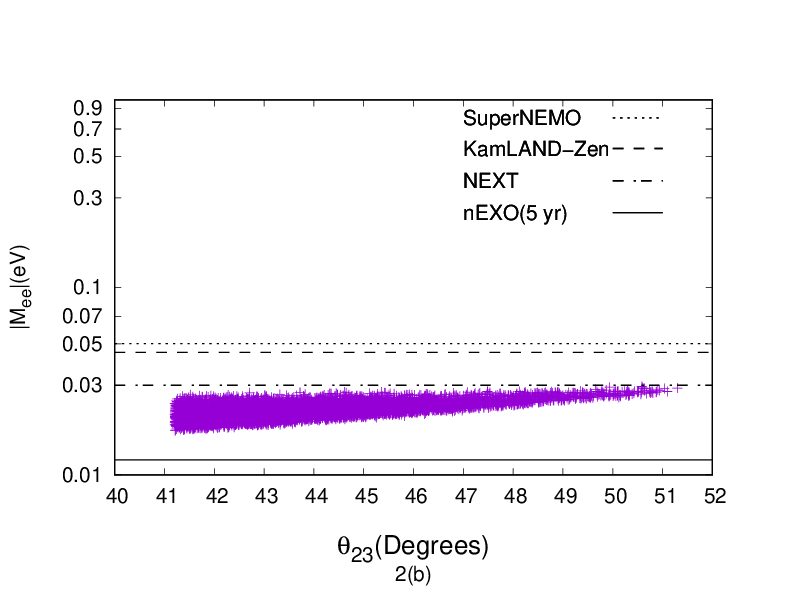}\\  
\end{tabular}
  \caption{\textbf{$b_1$-Case:} The $\theta_{23}-|M_{ee}|$ correlation plot for IH under LMA (left) and Dark-LMA (right) solutions.}
  \label{fig2}
\end{figure}

 \noindent Considering the neutrino oscillation data and range of scaling parameter $s$, the mass ratio $\frac{m_3}{m_1}$ is always less than 1. Thus, only IH is allowed in $b_1$ mass model. 


  \begin{figure}[H]
  \centering
{\includegraphics[width=0.45\linewidth]{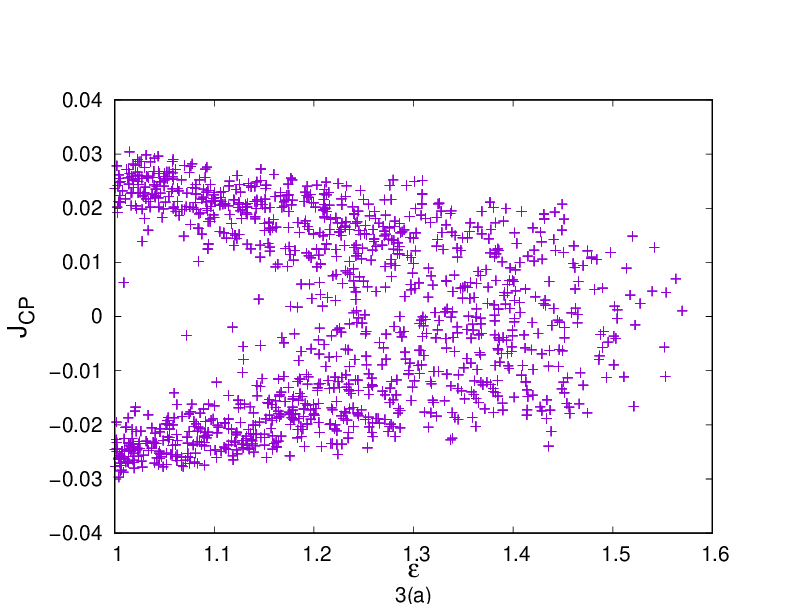} \includegraphics[width=0.45\linewidth]{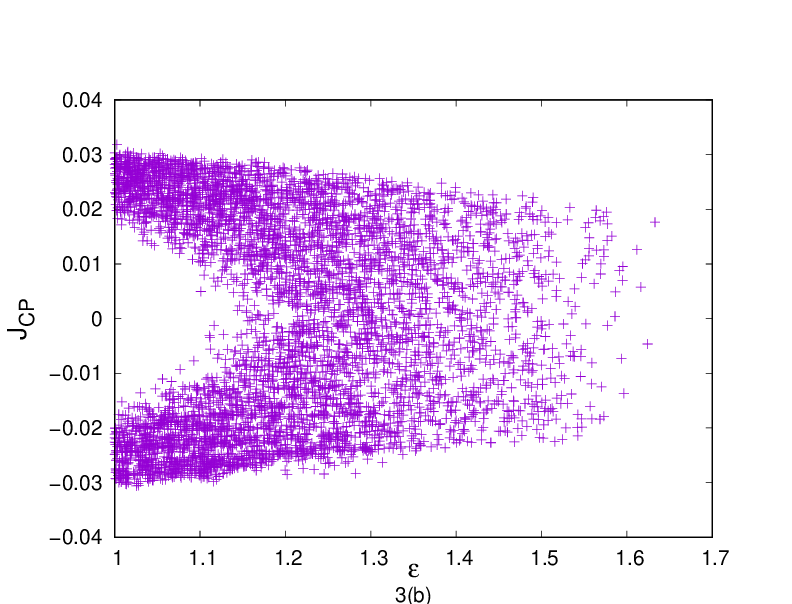}} 
{\includegraphics[width=0.45\linewidth]{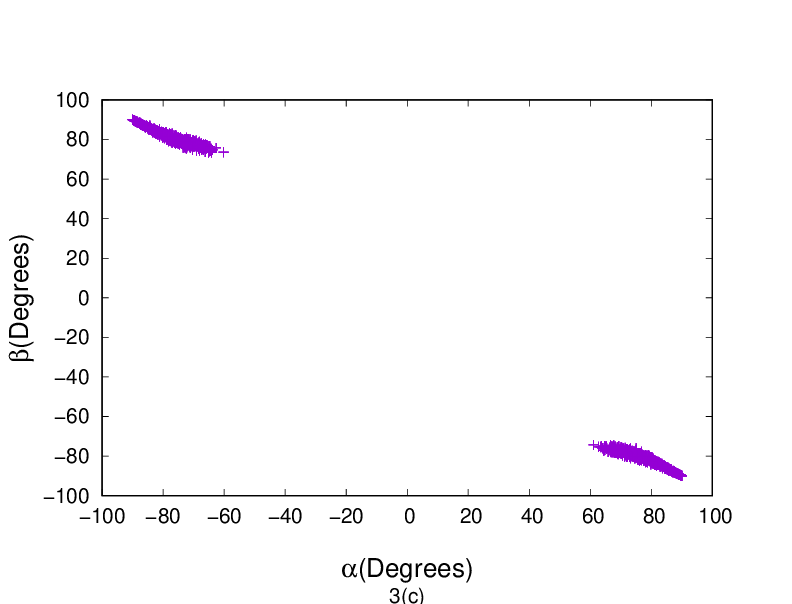} \includegraphics[width=0.45\linewidth]{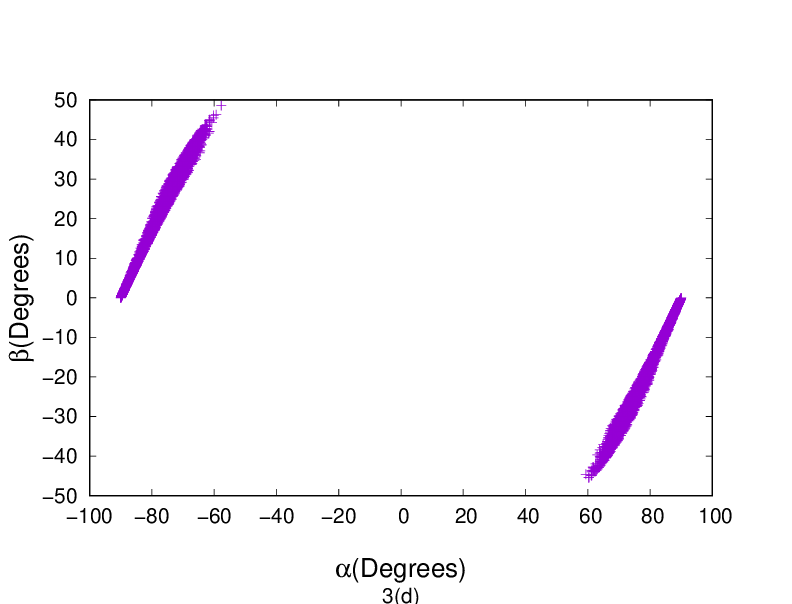}} 
 {\includegraphics[width=0.45\linewidth]{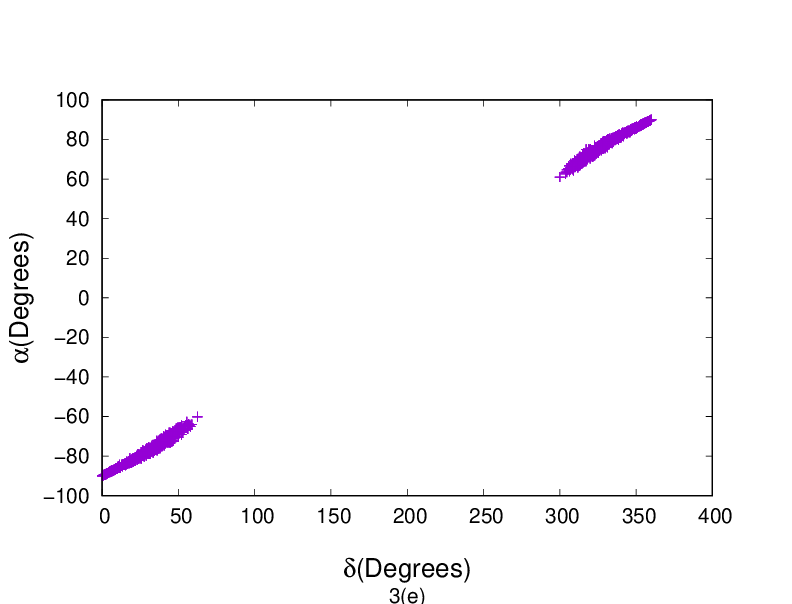} \includegraphics[width=0.45\linewidth]{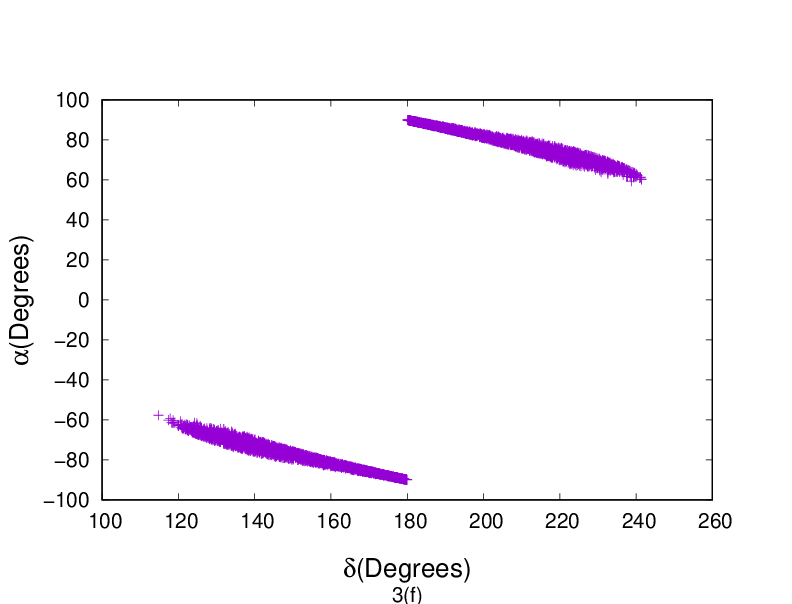}} 
  {\includegraphics[width=0.45\linewidth]{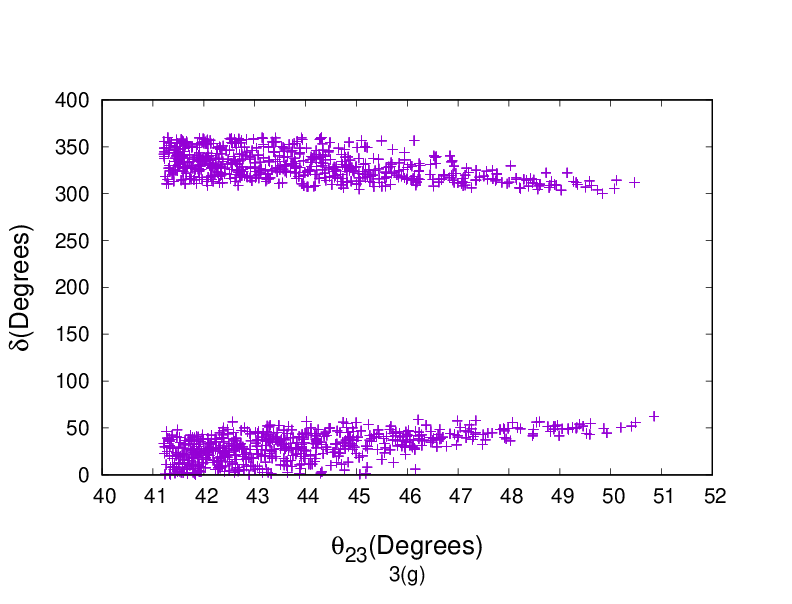} \includegraphics[width=0.45\linewidth]{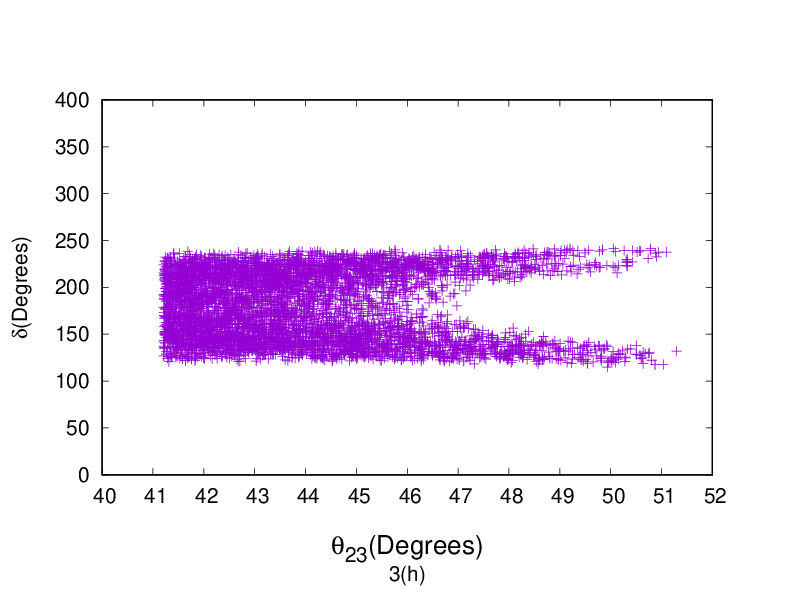}} 
  \caption{\textbf{$b_1$-Case:} The correlation plots under LMA (left panel) and Dark-LMA (right panel) solutions.}
  \label{fig3}
\end{figure}
 \noindent In Figs. \ref{fig1}-\ref{fig3}, we have depicted the predictions of the model as correlation plots amongst different parameters. The atmospheric mixing angle $\theta_{23}$ is allowed above and below maximality. The scale breaking parameter $\epsilon$ exhibit interesting correlation with $\theta_{23}$. For example, for $\theta_{23}\ge 45^o$, there exist an upper bound on $\epsilon\le 1.3$. The same bound in Dark-LMA scenario is $\approx 1.35$ (Figs. 1(a) and 1(b)).
 Also, for a given value of $\theta_{12}$, the upper bound on $\epsilon$ increases for LMA parameter space, however, it decreases (beyond $55.5^o$) for Dark-LMA (Figs. 1(c) and 1(d)). The precision measurement of NSI parameters dictating Dark-LMA parameter space shall have important consequences for the model as evident from Fig. 1(d). For example, if $\theta_{12}>56^o$ then the breaking in the model is allowed up to $\epsilon<1.5$.

 \noindent In Fig. \ref{fig2}, we have shown the correlation plots between $0\nu\beta\beta$ decay amplitude $|M_{ee}|$ and $\theta_{23}$. All the points shown in the figure are consistent with cosmological bound on sum of neutrino masses, $\sum m_i<0.12$ eV\cite{Planck:2018vyg}. The experimental sensitivities of $0\nu\beta\beta$-decay experiments\cite{nemo,zen,next1,next2,nexo} are also shown from which it is clear that nEXO\cite{nexo} has the required sensitivity to probe the model prediction for $M_{ee}$ in both LMA and Dark-LMA scenarios.

 \noindent One of the objectives of the broken scaling ansatz, analyzed in the present work, is to allow CP violation in model. In Fig. \ref{fig3}, we have shown the prediction of the model for CP violating phase. It is interesting to note that the model has distinguishing features for LMA and Dark-LMA as far as Majorana-type CP violation is concerned. The allowed parameter space, in $\epsilon-J_{CP}$ plane, shown in Fig. \ref{fig3}(a) can be, broadly, divided into three regions (i) $1\le\epsilon\le 1.2$ (ii) $1.2<\epsilon\le1.3$ and (iii) $\epsilon>1.3$. The region-I, is necessarily CP violating, however, in region-II and III both CP conserving and CP violating solutions are possible.  In view of Fig. \ref{fig1}(a), these $\epsilon$-regions can be translated in terms of $\theta_{23}$, for example, region-III ($\epsilon>1.3$) corresponds to $\theta_{23}<45^o$. Similar analysis can be done for Dark-LMA scenario using Fig. \ref{fig3}(b) where region-III ($\epsilon>1.35$) corresponds to $\theta_{23}<45^o$. The three CP violating phases $\delta,\alpha$ and $\beta$ exhibit strong correlation as depicted in Figs. \ref{fig3}(c-f). The Majorana CP phases $\alpha,\beta$ show positive correlation  whereas they are anti-correlated in Dark-LMA scenario (Figs. \ref{fig3}(c-d)). Similar is the prediction for $\delta-\alpha$ correlation for both LMA and Dark-LMA paradigms (Figs. \ref{fig3}(e-f)). Both LMA and Dark-LMA solutions predicts similar range of Majorana phase $\alpha$. However, LMA and Dark-LMA solutions have distinguishing implications for $\delta$ and $\beta$ CP phases (Table \ref{data1}).

 \noindent Figs. \ref{fig3}(g-h), describe correlation between $\theta_{23}$ and CP phase $\delta$ for LMA and Dark-LMA (which is important for CP violation studies). Note that for $\theta_{23}$ values near the upper $3\sigma$ bound (see Table \ref{data}), the CP is maximally violated ($\delta$ around $90^o$ or $270^o$) in the model which is interesting especially in light of the most precise measurement of $\delta$ (see Table \ref{data}) though definitive measurement of $\delta$ is still elusive.

 \begin{figure}[t]
 \centering
\begin{tabular}{c} 
\includegraphics[width=0.45\linewidth]{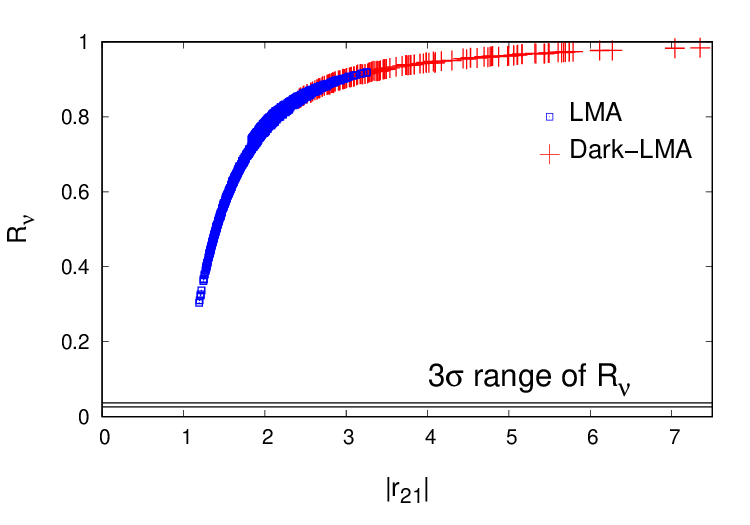}
\end{tabular}
\caption{\textbf{$b_2$-Case:} The correlation plots between  $|r_{21}|$ and $R_\nu$.}
  \label{figrnu}
\end{figure}

\begin{figure}[H]
  \centering
\begin{tabular}{cc} 
\includegraphics[width=0.45\linewidth]{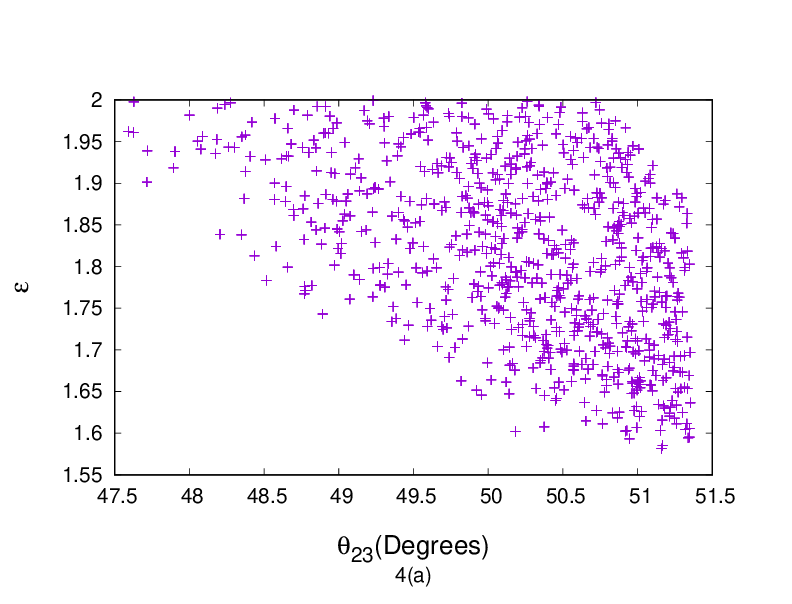}& \includegraphics[width=0.45\linewidth]{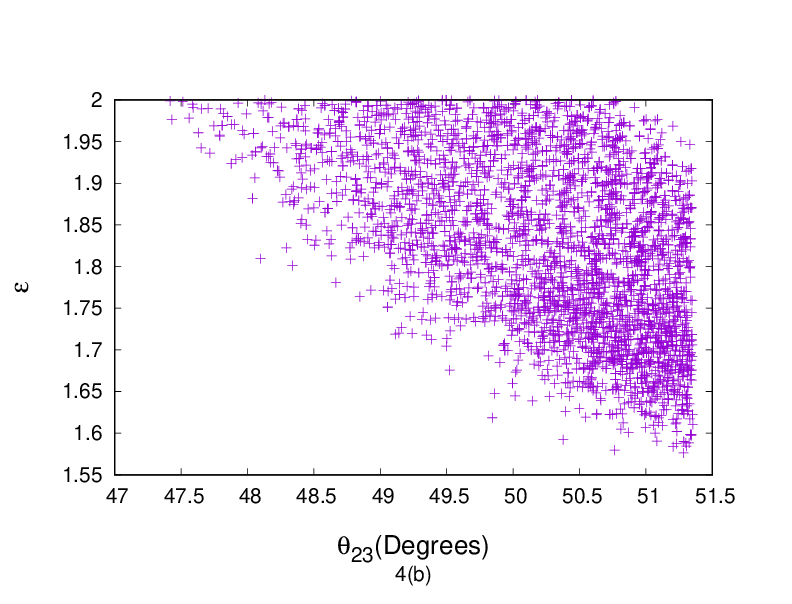}\\ 
 \includegraphics[width=0.45\linewidth]{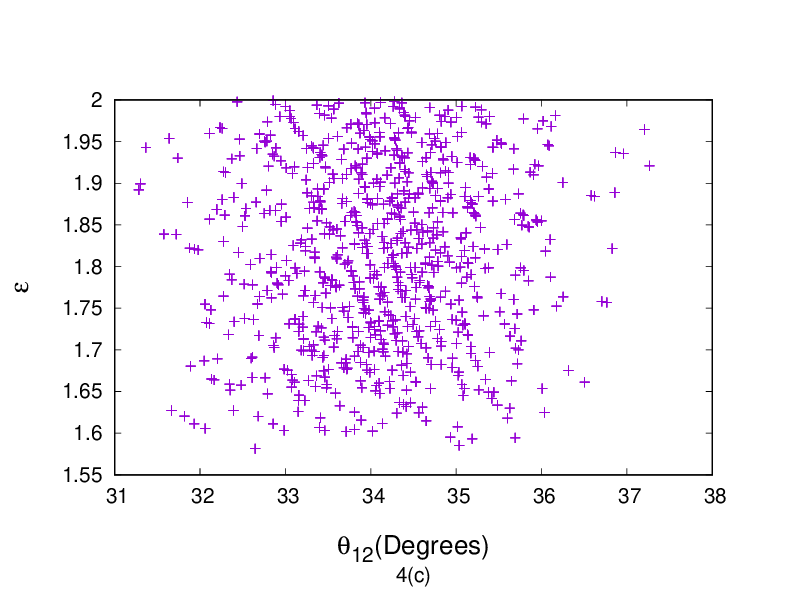}& \includegraphics[width=0.45\linewidth]{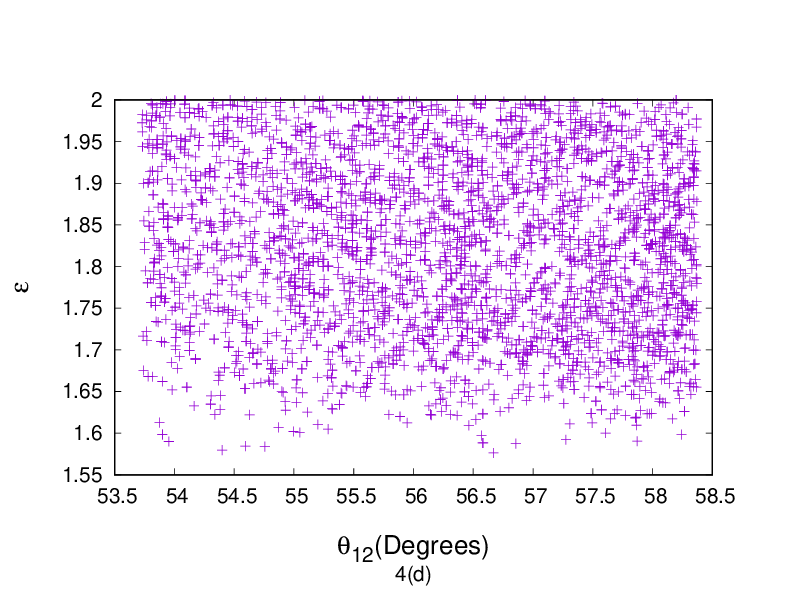}\\ 
\end{tabular}
  \caption{\textbf{$b_3$-Case:} The correlation plots between mixing angles $\theta_{12}$ and $\theta_{23}$ with scale breaking parameter $\epsilon$ with IH under LMA (left panel) and Dark-LMA (right panel) solutions.}
  \label{fig4}
\end{figure}

\subsection{$b_2$-Case}

The Taylor series expansion of the mass ratios up to the first order in $s_{13}$ for $b_2$ case is given by
\begin{eqnarray}
     &&|r_{21}|=\frac{m_2}{m_1} \approx 1 - \frac{ss_{13}  \cos\delta}{c_{12} s_{12} c_{23}  (c_{23} + ss_{23})},\\
      &&|r_{31}|=\frac{m_3}{m_1}\approx \frac{s_{23} (c_{23} + ss_{23})}{c_{23} (c_{23}s - s_{23})}-\frac{c_{12}s_{23} s_{13}\cos\delta}{c_{23}^2 s_{12} (s_{23} - s c_{23})}.
  \end{eqnarray} 
Using the mass ratios and neutrino oscillation data, we calculate $R_\nu$ defined in Eqn. (\ref{rnu}) for both LMA and Dark-LMA scenarios. The results have been shown in Fig. \ref{figrnu} which clearly ruled out this case as $R_{\nu}$ is well outside its $3\sigma$ allowed experimental range. Thus, broken scaling, to explain non-zero $\theta_{13}$ and CP violation, cannot be achieved \textit{via} second row of the scaled neutrino mass matrix.

\begin{figure}[t]
  \centering
\begin{tabular}{cc} 
\includegraphics[width=0.45\linewidth]{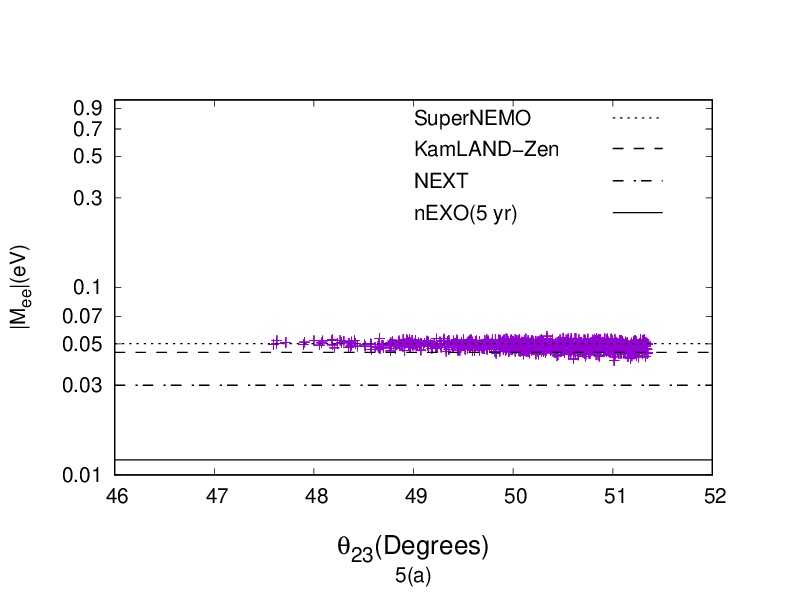}& \includegraphics[width=0.45\linewidth]{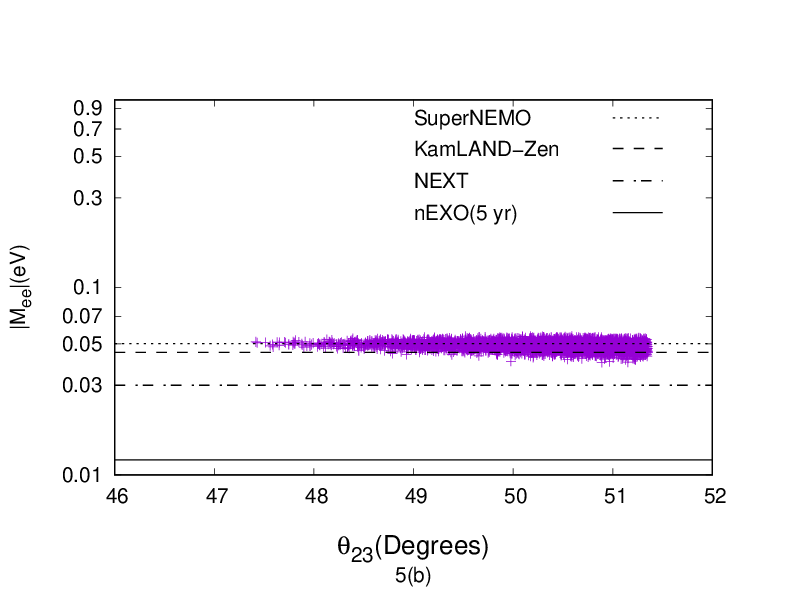}\\ 
 \includegraphics[width=0.45\linewidth]{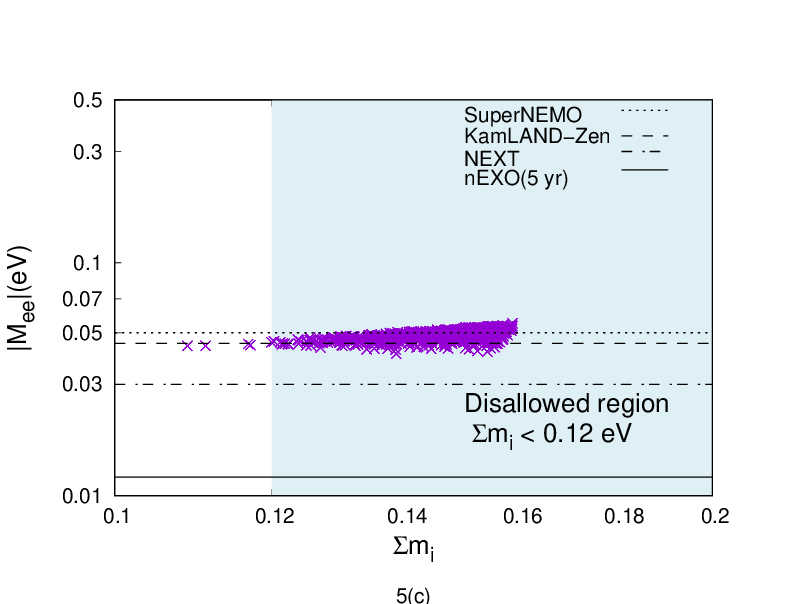}& \includegraphics[width=0.45\linewidth]{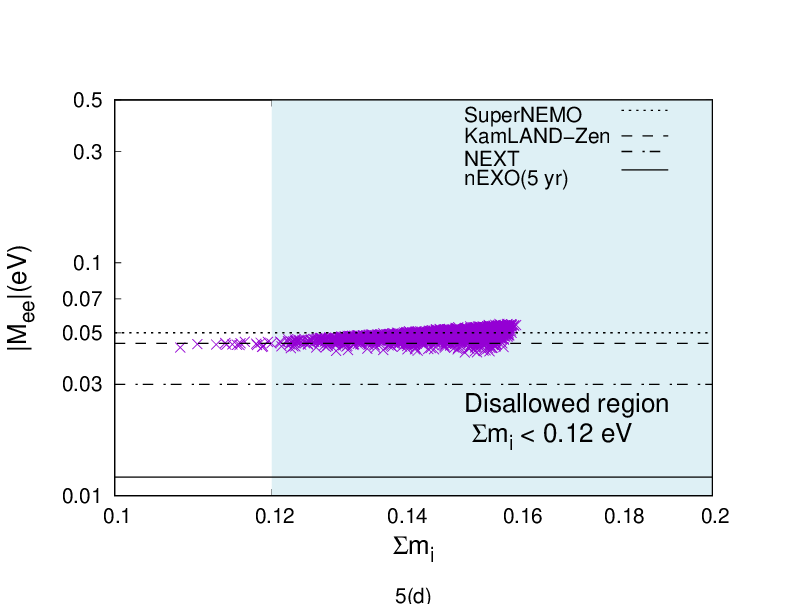}\\ 
\end{tabular}
  \caption{\textbf{$b_3$-Case:} The $\left(\theta_{23},\sum m_i\right)-|M_{ee}|$ correlation plots for IH under LMA (left panel) and Dark-LMA (right panel) solutions. The shaded region shows parameter space disallowed by cosmological bound, $\sum m_i<0.12$ eV.}
  \label{fig5}
\end{figure}

\begin{figure}[H]
  \centering
\begin{tabular}{cc} 
\includegraphics[width=0.45\linewidth]{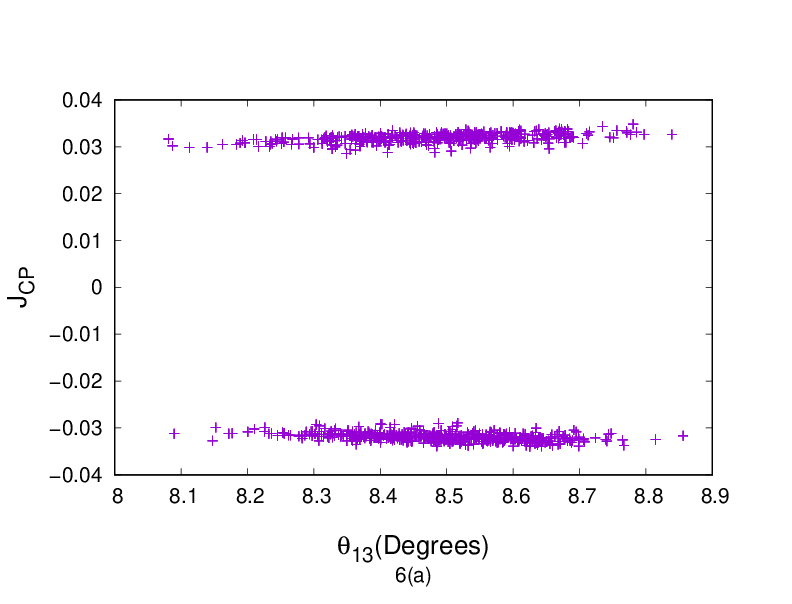}& \includegraphics[width=0.45\linewidth]{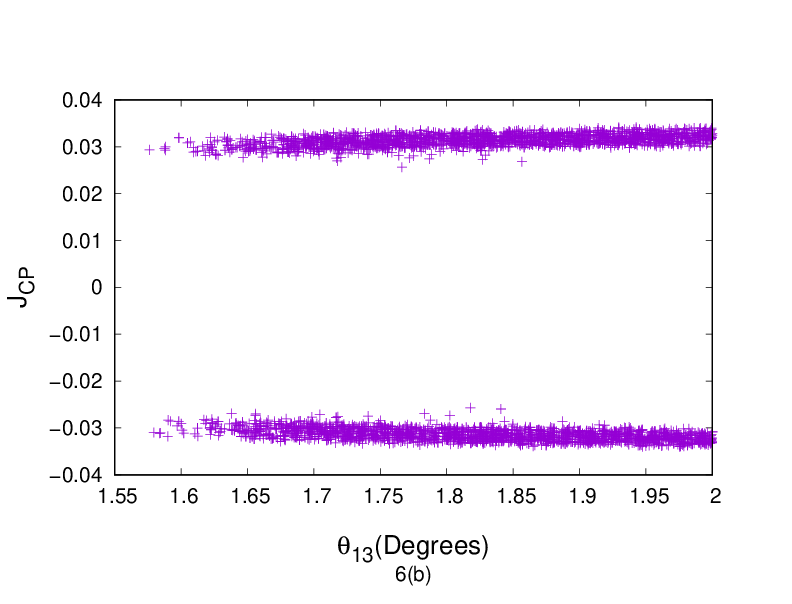}\\ 
 \includegraphics[width=0.45\linewidth]{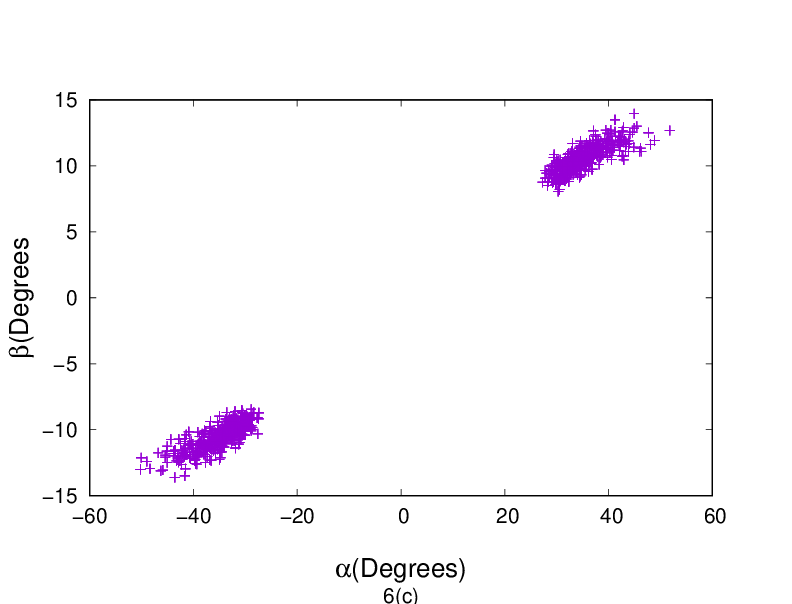}& \includegraphics[width=0.45\linewidth]{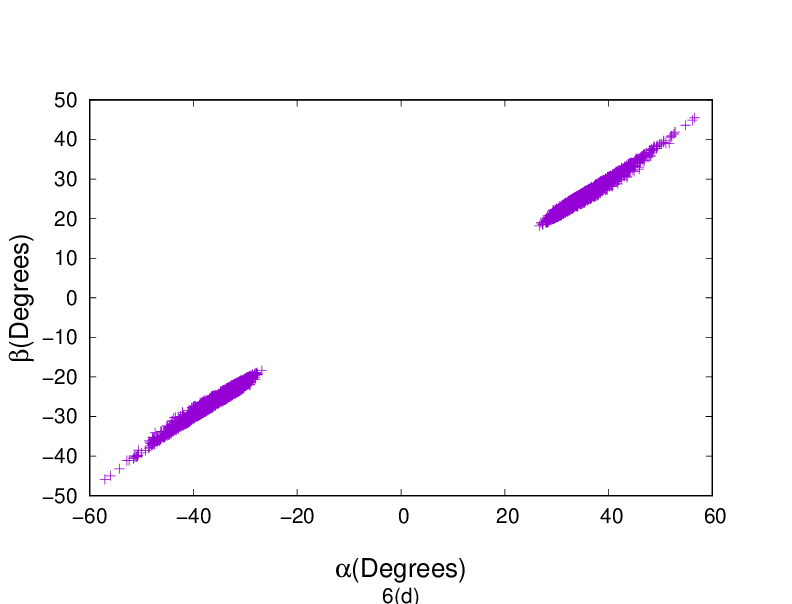}\\ 
 \includegraphics[width=0.45\linewidth]{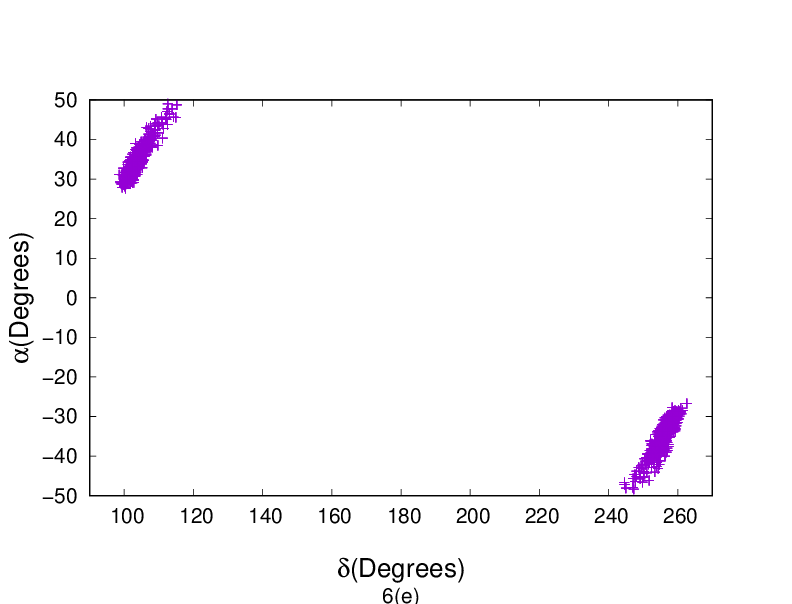}& \includegraphics[width=0.45\linewidth]{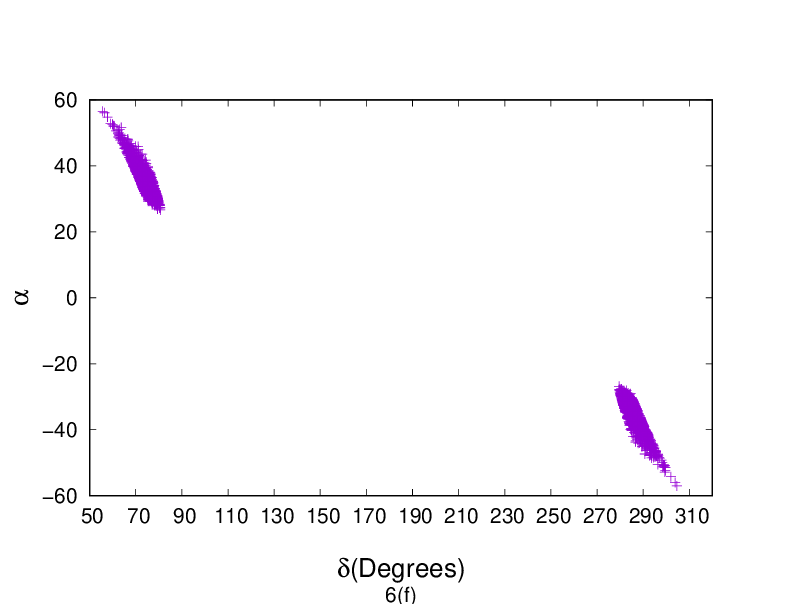}\\ 
  \\ \end{tabular}
  \caption{\textbf{$b_3$-Case:} The correlation plots under LMA (left panel) and Dark-LMA (right panel) solutions.}
  \label{fig6}
\end{figure}

\subsection{$b_3$-Case}
For this case, the  the Taylor series expansion of mass ratios up to first order in $s_{13}$ is given by
\begin{eqnarray}
 && |r_{21}|=\frac{m_2}{m_1} \approx  1 + \frac{s_{13}  \cos\delta}{c_{12} s_{12} s_{23} (c_{23} + s~ s_{23})},\\
 &&   |r_{31}|=\frac{m_3}{m_1}\approx \frac{c_{23} (c_{23} + ss_{23})}{s_{23} (c_{23}s - s_{23})}-\frac{c_{12}c_{23} s_{13}\cos\delta}{s_{23}^2 s_{12} (s_{23} - s c_{23})}.
\end{eqnarray}

\noindent The correlation plots for this case are given in Figs. \ref{fig4}-\ref{fig6}. The correlations amongst the parameters are different to that of $b_1$-case The atmospheric mixing angle $\theta_{23}$ is restricted to have above maximal values only. Thus, in order to have non-zero $\theta_{13}$ and to allow CP violation in scaled  neutrino mass matrix, $\theta_{23}$ must be above $45^o$. In fact, larger the $\theta_{23}$-deviation from $45^o$, more wider range of $\epsilon$ is allowed by current data.

\noindent In Fig. \ref{fig5}, the correlation plots ($\theta_{23}-|M_{ee}|$) and ($\sum m_i-|M_{ee}|$) are shown. The lower bound on $|M_{ee}|$ is larger as compared to $b_1$-case. The predicted bound $|M_{ee}|>0.04$ eV is reachable at NEXT\cite{next1,next2} and nEXO\cite{nexo} experiments as their sensitivities are high. Also, cosmological observations revealing $\sum m_i$ may have important consequence on the viability of the broken scaling considered here. For example, the more stringent bound on $\sum m_i$, coming from inclusion of the baryon acoustic oscillation (BAO) analysis by Dark Energy Spectroscopic Instrument (DESI) data \cite{DESI:2024mwx}, of $\sum m_i<0.072$ eV (at 95\% C. L.) will rule out both $b_1$ and $b_3$ scenarios. Furthermore, this case is found to be necessarily CP violating as only non-zero values of Jarlskog parameter ($J_{CP}$), $\alpha$ and $\beta$ are allowed (Fig. \ref{fig6}). The CP violating phases are strongly constrained in the ranges given in fourth column of Table \ref{data1}.

\section{Conclusions}\label{sec4}
The breaking of scaling in $M_\nu$ is required to explain non-zero value of $\theta_{13}$, deviation of $\theta_{23}$ from maximality and to allow CP violation in the leptonic sector. In general, theoretical models to understand emerged picture of neutrino masses and mixing matrices, often, are based on introducing additional degrees of freedom to particle content of the SM resulting in new couplings which may result in non-standard interactions (NSIs) of neutrino, thus, obscuring structure of the neutrino mixing matrix at sub-leading level. The Dark-LMA solution is the culmination of such NSIs giving new solution space consistent with neutrino oscillation data with $\sin^2\theta_{12}\simeq 0.7$. We have investigated three patterns ($b_1,b_2 ~\text{and}~ b_3$) of broken ``scaling" ansatz of the neutrino mass matrix under the paradigms of the LMA and Dark-LMA solutions of the neutrino oscillations.

\noindent The pattern $b_2$ is disallowed as it is not in consonance with $3\sigma$ experimental range of $R_\nu$. The cases $b_1$ and $b_3$ exhibits distinguishing phenomenology. For example, while in $b_1$, $\theta_{23}$ is found to be above as well as below maximality, only above maximal $\theta_{23}$-values accommodates LMA and Dark-LMA solutions in case of $b_3$. The lower bound on $|M_{ee}|$ is $\approx 0.015$ eV ($\approx0.045$ eV) in $b_1$ ($b_3$) which can be probed in nEXO (NEXT and nEXO) $0\nu\beta\beta$ decay experiment(s). The non-observation of this process and more stringent bound on $\sum m_i$, coming from inclusion of the baryon acoustic oscillation (BAO) analysis by Dark Energy Spectroscopic Instrument (DESI) data, of $\sum m_i<0.072$ eV (at 95\% C. L.) will rule out both $b_1$ and $b_3$ scenarios.

\begin{table}[t]\label{data1}
\begin{center}
\begin{tabular}{c|l|l|l}
\hline 
CP Phase &Solution & $b_1$-Case & $b_3$-Case \\ \hline\hline
\multirow{2}{*}{$\alpha$} & LMA & ($-90^o\to-60^o$)$\oplus$($60^o\to90^o$) &  ($-50^o\to-23^o$)$\oplus$($27^o\to50^o$) \\ 
 & Dark-LMA & ($-90^o\to-60^o$)$\oplus$($60^o\to90^o$) & ($-55^o\to-28^o$)$\oplus$($28^o\to56^o$) \\ \hline

 \multirow{2}{*}{$\beta$} & LMA &($70^o\to90^o$)$\oplus$($-90^o\to-70^o$) & ($-13^o\to-8^o$)$\oplus$($8^o\to14^o$)  \\ 
 & Dark-LMA & ($0^o\to45^o$)$\oplus$($-45^o\to0^o$) & ($-46^o\to-18^o$)$\oplus$($18^o\to46^o$) \\ \hline

 \multirow{2}{*}{$\delta$} & LMA & ($0^o\to55^o$)$\oplus$($300^o\to360^o$) & ($90^o\to116^o$)$\oplus$($244^o\to262^o$)  \\ 
 & Dark-LMA & ($115^o\to180^o$)$\oplus$($180^o\to240^o$) & ($55^o\to85^o$)$\oplus$($280^o\to305^o$) \\ \hline

\end{tabular}
\caption{\label{data1} The $3\sigma$ range of the CP violating phases.}
\end{center}
\end{table}

\noindent One of the features of the scale breaking patterns is the predictions for CP violating parameters. The $b_3$-case is found to be necessarily CP violating while $b_1$ can have both CP conserving and violating solution spaces. In fact, $b_1$ is, also, necessarily CP violating for $1\le\epsilon\le1.2$ with LMA solution. There exist an interesting interplay of $\theta_{23}$-octant and possible CP violation in the model. For example, in $b_1$, for $\theta_{23}<45^o$ both CP conserving and violating solutions are allowed. However, for LMA, as $\theta_{23}$ take values near the upper end of the allowed $3\sigma$ range (i.e. $\approx 50^o$) $\delta$ approaches its maximal value of $90^o$ or $270^o$. Thus, in future, if $\theta_{23}$ is confirmed to lie around these values (as indicated by current data, see Table \ref{data}) it will, also, be a confirmation of existence of CP violation in the leptonic sector.

\noindent The CP violating phases are constrained in narrow ranges as shown in Table \ref{data}. The current data on Dirac-type CP phase $\delta$ (seventh row of third column in  IH part of Table \ref{data}) disallows first region of $\delta$ for both LMA and Dark-LMA solutions in $b_1$ and $b_3$ cases. Also, the correlations of Majorana phases ($\alpha,\beta$) can be probed in $0\nu\beta\beta$ decay experiments. It is interesting since the model-predicted ($\alpha-\beta$) correlation has the capability of distinguishing LMA and Dark-LMA solutions. For example, in $b_1$-case, ($\alpha-\beta$) correlation coefficient is negative (positive) for LMA (Dark-LMA) solution, however, such distinction is not possible in $b_3$-case. In general, LMA and Dark-LMA solutions have distinguishing implication for CP violating phases ($\beta,\delta$) as is evident from Table \ref{data1}.

\noindent\textbf{\Large{Acknowledgments}}
 \vspace{.3cm}\\
AK acknowledges Central University of Himachal Pradesh (CUHP) for providing financial assistance in the form of freeship. DG acknowledges Department of Science and Technology (DST) for providing financial assistance through DST-INSPIRE programme vide No. 202000015819. The authors, also, acknowledge Department of Physics and Astronomical Science for providing necessary facility to carry out this work.


\end{document}